\documentclass[journal]{IEEEtran}
\usepackage{amsmath,amsfonts}
\usepackage{algorithmic}
\usepackage{array}
\usepackage[caption=false,font=footnotesize]{subfig}
\usepackage{textcomp}
\usepackage{stfloats}
\usepackage{url}
\usepackage{verbatim}
\usepackage[pdftex]{graphicx}
\usepackage[colorlinks=true,linkcolor=blue,citecolor=blue,urlcolor=blue]{hyperref}

\usepackage{balance}

\begin{document}
\title{
Mitigation of Structural Harmonic Instability in Virtual Admittance-Based Grid-Forming Inverters via Mimicking Skin Effect
}
\author{
Jaekeun~Lee,~\IEEEmembership{Graduate~Student~Member,~IEEE,}
Jae-Jung~Jung,~\IEEEmembership{Senior~Member,~IEEE,}
Shenghui~Cui,~\IEEEmembership{Senior~Member,~IEEE}
}

\markboth{IEEE Transactions on power electronics}{}

\maketitle

\begin{abstract}
The virtual admittance-current controller (VA-CC) scheme is widely employed
to emulate an equivalent inductance in front of the internal voltage source of grid-forming inverters.
However, recent studies have reported harmonic instabilities associated with VA-CC,
motivating the need for a more physically interpretable understanding of their origin.
This letter identifies a delay-independent structural mechanism of
harmonic instability in the VA-CC scheme,
wherein the interaction between the filter and virtual inductances introduces a non-passive second-order transfer-function term exhibiting negative resistance.
To address this issue, a simple yet effective modification is proposed by
integrating a parallel virtual resistor into the VA structure.
This reconfiguration enhances the passivity of VA-CC scheme across the harmonic range by mimicking the skin effect which augments damping in high-frequency range,
without altering the well-established current controller or voltage feedforward control.
Experimental results validate that the proposed method achieves robust harmonic stability, whereas the conventional approach fails under identical grid conditions.
\end{abstract}


\IEEEpeerreviewmaketitle

\section{Introduction}
\IEEEPARstart{W}{ith} the rapid transition toward high-penetration renewable energy systems,
grid-forming (GFM) inverters have become indispensable for ensuring the stability and
resilience of modern power grids \cite{cite0_Rosso2021GFM_Review}. To enhance the stability of GFM control under
various grid conditions, virtual impedance-based control has been widely adopted,
with virtual admittance (VA) being particularly preferred due to its implementation
that avoids noise-sensitive differentiators \cite{cite1_Rodriguez2013VA}.
Furthermore, by incorporating a dedicated current controller (CC),
the VA-CC structure provides robust current regulation—a feature that is vital for practical industrial applications.
In this structure, voltage feedforward control (VFF) is traditionally regarded as an essential element
to enhance current control dynamics and is a well-established standard in conventional grid-connected inverters.
Therefore, this letter focuses on the VA-CC defined as the integration of
VA, CC, and VFF, to preserve the high-performance standards of conventional power electronics.

However, prior works have revealed the instability issues
regarding the VA-CC, and suggest to modify CC or VFF to enhance the
stability of VA-CC \cite{cite2_Gao2026VA_MMC, cite3_Obi2025RevaluationVA}.
While effective for stabilization, these approaches inherently compromises the
current regulation performance and degrades the control dynamics.
Such a trade-off is often undesirable in practical industrial applications,
as it deviates from the well-established current control standards.
Consequently, there remains a need for a stability enhancement strategy
that preserves the conventional CC and VFF structures.

This letter provides a simple insight into the instability mechanism of the VA-CC,
and proposes a simple yet effective VA-level solution that ensures stability without modifying CC or VFF.
Specifically, the interaction between the VA and the filter inductance introduces a
non-passive second-order term into the inverter output impedance,
which becomes particularly problematic in the harmonic frequency range.
While emulating a series \(R\!-\!L\) structure has long been the standard for VA implementation,
this structure inherently lacks the high-frequency damping capability
needed to neutralize such non-passivity.
To address this, this letter proposes a modified VA design
incorporating a virtual resistor in parallel with the virtual inductor.
By emulating the resistive behavior of physical inductors at high frequencies (i.e., skin effect),
the proposed parallel virtual resistor ensures VA to be resistive in the harmonic range,
thereby extensively enhancing the passivity of the VA-CC system.
Experimental results are provided to validate the effectiveness of the proposed method.

\section{Analysis on the Instability of VA-CC}

The investigated GFM control scheme comprises a VA, CC,
and VFF, where the outer loop emulates an internal voltage
source (IVS), denoted by \(\mathbf{e}\).
The voltage at the point of common coupling (PCC) and the inverter output current are
denoted by \(\mathbf{v}_\text{pcc}\) and \(\mathbf{i}\), respectively.
Throughout this letter, the boldface symbols represent complex vectors in the
stationary reference frame.
Notably, control delays are intentionally neglected specifically in this section.
This is to demonstrate that the non-passive behavior is not merely an artifact
of control delays, but rather an inherent characteristic originating from the
VA-CC structure itself.

\subsection{Formulation of Equivalent Output Impedance}

The current reference \(\mathbf{i}_\text{ref}\) is generated by the VA as
\(\mathbf{i}_\text{ref} = Y_\text{v}(s) (\mathbf{e} - \mathbf{v}_\text{pcc})\),
where \( Y_\text{v}(s) \) denotes the VA transfer function.
The CC, \( G_\text{cc}(s) \), is implemented using a proportional-resonant (PR) structure
in the stationary reference frame \cite{cite4_Baeckeland2022GFM_FL}.
Notably, in the harmonic frequency range,
\(G_\text{cc}(s)\) is predominantly governed by its proportional gain
\(K_\text{cc,p}\), regardless of whether a PR controller or a \(dq\)-frame PI controller
is employed. This ensures that the following stability analysis remains generalized for
standard current control frameworks.

By incorporating the VFF,
the output voltage of the inverter is expressed as:
\begin{equation}
  \mathbf{v}_\text{o}
  = G_\text{cc}(s) Y_\text{v}(s) \left( \mathbf{e} - \mathbf{v}_\text{pcc} \right)
  - G_\text{cc}(s) \mathbf{i}
  + \mathbf{v}_\text{pcc},
  \label{eq:1}
\end{equation}
where \(\mathbf{v}_\text{o}\) is the PWM output voltage.
Considering a filter inductance \(L_\text{f}\), the relationship
between \(\mathbf{v}_\text{o}\) and \(\mathbf{v}_\text{pcc}\) is given by
\(\mathbf{v}_\text{o} = sL_\text{f} \mathbf{i} + \mathbf{v}_\text{pcc}\).
Substituting this into \eqref{eq:1} yields:
\begin{equation}
  \mathbf{v}_\text{pcc} = \mathbf{e}
  - \left( \frac{1}{Y_\text{v}(s)} + \frac{sL_\text{f}}{G_\text{cc}(s) Y_\text{v}(s)} \right)\mathbf{i}.
  \label{eq:2}
\end{equation}
In the harmonic range, where the resonant gain of the CC is negligible,
\(G_\text{cc}(s) \approx K_\text{cc,p}\).
Consequently, the IVS appears behind an equivalent output impedance \(Z_\text{eq}(s)\),
defined as:
\begin{equation}
  Z_\text{eq}(s) = \frac{1}{Y_\text{v}(s)} + \frac{sL_\text{f}}{G_\text{cc}(s) Y_\text{v}(s)} 
  \approx \frac{1}{Y_\text{v}(s)} + \frac{sL_\text{f}}{K_\text{cc,p} Y_\text{v}(s)}.
  \label{eq:3}
\end{equation}
Ideally, if the CC gain were infinite \( \left( G_\text{cc}(s) \rightarrow \infty \right) \) across all frequencies,
the inverter would perfectly track the reference, \(\mathbf{i} = \mathbf{i}_\text{ref}\).
In practice, however, the control bandwidth of CC is inherently limited by the control delay,
and its magnitude is limited to the proportional gain \(K_\text{cc,p}\) in the harmonic frequency range.
This finite control gain introduces an unintended term, \({sL_\text{f}}/(G_\text{cc}(s)Y_\text{v} (s))\),
into \(Z_\text{eq}(s)\) alongside the desired virtual impedance \(1 / Y_\text{v} (s)\).

\begin{figure}[t]
  \centering
  \subfloat[]{
    \includegraphics[width=0.50\linewidth,keepaspectratio,
    trim=0mm 0mm 0mm 0mm,clip]{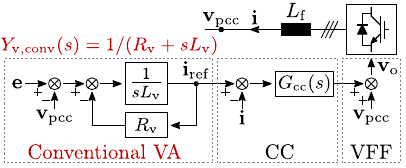}
  }
  \vspace{-15pt}
  \subfloat[]{
    \includegraphics[width=0.40\linewidth,keepaspectratio,
    trim=0mm 0mm 0mm 0mm,clip]{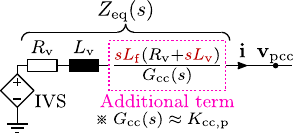}
  }
  \subfloat[]{
    \includegraphics[width=0.45\linewidth,keepaspectratio,
    trim=0mm 0mm 0mm 0mm,clip]{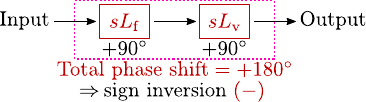}
  }
  \caption{
  VA-CC instability: (a) control structure, (b) equivalent circuit of \(Z_\text{eq}(s)\),
  and (c) emergence of the negative resistance term, \(s^2 L_\text{v}L_\text{f}\).}
  \label{fig:ConventionalDiagrams}
\end{figure}

\subsection{Inherent Non-passivity of Conventional VA-CC}

Conventionally, the VA is designed based on a series \(R\!-\!L\) structure,
denoted by
\(Y_\text{v,conv}(s) = (R_\text{v} + sL_\text{v})^{-1}\).
Substituting this into \eqref{eq:3} yields:
\begin{equation}
  \begin{aligned}
  Z_\text{eq}(s) 
  & = {R_\text{v} + s L_\text{v}} + \frac{sL_\text{f} (R_\text{v} + s L_\text{v}) }{G_\text{cc}(s)}
  \\
  & \approx {R_\text{v} + s L_\text{v}} + \frac{sL_\text{f} (R_\text{v} + s L_\text{v}) }{K_\text{cc,p}}.
  \end{aligned}
  \label{eq:5}
\end{equation}
The equivalent impedance includes a second-order term, \(s^2 L_\text{f}L_\text{v}/K_\text{cc,p}\).
For any \(s=j\omega\), this term simplifies to a negative resistance,
\(-\omega^2 L_\text{f} L_\text{v}/K_\text{cc,p}\).
In the harmonic frequency range, this negative resistance becomes dominant,
thereby compromising the passivity and harmonic stability of VA-CC scheme.

Fig.~\ref{fig:ConventionalDiagrams} illustrates the vulnerability of the conventional VA-CC.
The control structure and its equivalent circuit are shown in
Fig.~\ref{fig:ConventionalDiagrams}(a) and (b), respectively. As depicted in
Fig.~\ref{fig:ConventionalDiagrams}(c), the interaction between \(sL_\text{f}\) and \(sL_\text{v}\)
introduces a second-order term that causes \(180^\circ\) phase shift (i.e., sign inversion).
This characteristic effectively acts as a negative resistance,
rendering the VA-CC non-passive in the harmonic range.

\section{Proposed VA with Parallel Virtual Resistor}

In the harmonic range, the product of two inductive terms, \(sL_\text{f}\) and \(sL_\text{v}\)
becomes dominant.
To mitigate the non-passivity arising from the \(s^2 L_\text{f} L_\text{v}\) term,
the proposed method reconfigures the VA structure.
A fundamental design challenge exists in the conflicting frequency-dependent requirements:
the VA must remain inductive at the fundamental frequency to fulfill GFM capability \cite{cite5_Wu2024CLC_GFM},
yet this inductive characteristic should be avoided in the harmonic range to prevent instability.
To bridge this gap, a parallel virtual resistor configuration is proposed,
as detailed in the following subsections.

\subsection{Physical Rationale and Design of the Proposed VA}

While a series \(R\!-\!L\) circuit is the standard model for the conventional VA schemes,
it assumes a constant resistance across all frequencies.
In reality, physical inductors exhibit frequency-dependent resistive losses,
such as skin effect, which become significant at higher frequencies.
This behavior can be interpreted as the presence of parallel dissipative circuit elements \cite{cite6_Kim1996SkinEffect, cite7_Matsumori2018LCL_ESR},
which increase the effective resistive component with frequency.
Inspired by this physical behavior, rather than strictly adhering to a constant series \(R\!-\!L\) structure,
this letter proposes a parallel virtual resistor configuration.
This approach provides a solution that satisfies
the inductive requirements at the fundamental frequency while ensuring
passivity in the harmonic range.

\begin{figure}[t]
  \centering
  \subfloat[]{
    \includegraphics[width=0.4\linewidth,keepaspectratio,
    trim=0mm 0mm 0mm 0mm,clip]{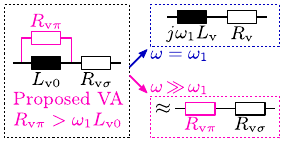}
  }
  \subfloat[]{
    \includegraphics[width=0.4\linewidth,keepaspectratio,
    trim=0mm 0mm 0mm 0mm,clip]{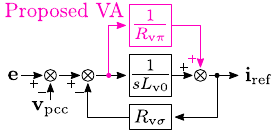}
  }
  \caption{
  Proposed VA structure: (a) circuit model of the proposed VA and (b) control implementation of the proposed VA.}
  \label{fig:ProposedDiagrams}
\end{figure}

For the proposed method, the virtual admittance is modeled as shown in
Fig.~\ref{fig:ProposedDiagrams}(a),
with its implementation depicted in Fig.~\ref{fig:ProposedDiagrams}(b).
Here, \(L_\text{v0}\), \(R_{\text{v}\pi}\), and \(R_{\text{v}\sigma}\) denotes
virtual inductance, virtual parallel resistance, and virtual series resistance of the proposed VA,
respectively.
The transfer function of the proposed VA, \(Y_\text{v,prop}(s)\) is given by:
\begin{equation}
Y_\text{v,prop}(s)
= \left( R_{\text{v}\sigma}
+ \frac{s R_{\text{v}\pi} L_\text{v0}}{R_{\text{v}\pi} + s L_\text{v0}} \right)^{-1}.
\end{equation}
By calculating the parameters \(L_\text{v0}\), \(R_{\text{v}\pi}\), and \(R_{\text{v}\sigma}\) as:
\begin{equation}
  L_\text{v0} = \left({X_\text{v}^2 + (R_\text{v} - R_{\text{v}\sigma})^2}\right)\left({\omega_1 X_\text{v}}\right)^{-1},
\end{equation}
\begin{equation}
  R_{\text{v}\pi} = \left({X_\text{v}^2 + (R_\text{v} - R_{\text{v}\sigma})^2}\right) \left({R_\text{v} - R_{\text{v}\sigma}}\right)^{-1},
\end{equation}
where \(R_{\text{v}\sigma} < R_\text{v}\),
the proposed VA ensures
\begin{equation}
  Y_\text{v,prop}(j\omega_1)
  = \left( R_\text{v} + j \omega_1 L_\text{v} \right)^{-1}.
\end{equation}
This ensures that the desired virtual resistance and inductance are preserved as
\(R_\text{v}\) and \(L_\text{v}\) respectively, at the fundamental frequency.

Meanwhile, in the harmonic frequency range \( \left( \omega \gg \omega_1 \right) \),
\(Y_\text{v,prop}(j\omega)\) is approximated as:
\begin{equation}
Y_\text{v,prop}(j\omega)
= \left( R_{\text{v}\sigma}
+ \frac{j\omega R_{\text{v}\pi} L_\text{v0}}{R_{\text{v}\pi} + j\omega L_\text{v0}} \right)^{-1}
\approx
\left( R_{\text{v}\sigma} + R_{\text{v}\pi} \right)^{-1}.
\label{eq:10}
\end{equation}
Substituting \eqref{eq:10} into the expression for \(Z_\text{eq}(s)\),
the problematic term \(sL_\text{f}/(G_\text{cc}(s) Y_\text{v}(s))\) becomes approximately
\(sL_\text{f}(R_{\text{v}\sigma} + R_{\text{v}\pi})/K_\text{cc,p}\).
Compared with the conventional method,
this reconfiguration effectively precludes the phase inversion by \(s^2\),
thereby enhancing the passivity of
\(Z_\text{eq}(s)\) in the harmonic range.
Furthermore, this approach also contributes to passivity through the first term,
\(1/Y_\text{v,prop}(s)\).
This is because the real part of \(1/Y_\text{v,prop}(s)\) increases with frequency,
in contrast to the conventional \(1/Y_\text{v,conv}(s)\),
whose real part remains equal to \(R_\text{v}\) over entire frequency range.

\subsection{Comparison of Conventional and Proposed VA-CC Schemes}

\begin{figure}[t]
  \centering
  \includegraphics[width=0.8\linewidth,keepaspectratio,trim=0mm 0mm 0mm 0mm,clip]{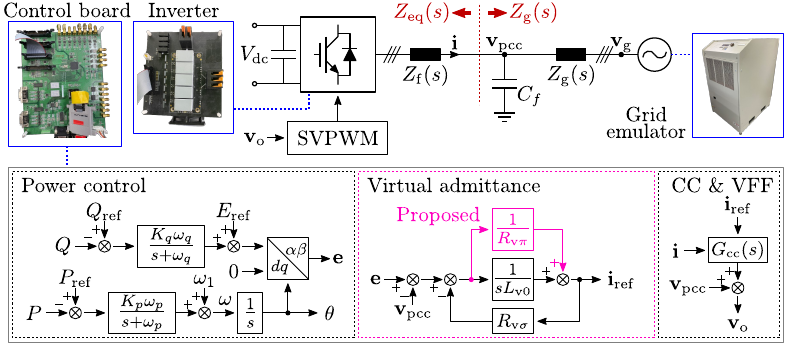}
  \caption{Experimental setup and control structure.}
  \label{fig:Setup}
\end{figure}

\begin{table}[t]
\centering
\caption{Experimental Parameters}
\label{tab:parameters}
\begin{tabular}{c c c}
\hline
\text{Parameter} & \text{Symbol} & \text{Value} \\
\hline
Rated power & \(P_\text{rated}\) & 3~kW \\
Fundamental grid frequency & \(\omega_1\) & \(2\pi \times 60\)~rad/s \\
Switching frequency & \(f_\text{sw}\) & 10~kHz \\
Sampling frequency & \(f_\text{s}\) & 20~kHz \\
Grid voltage (line-to-line RMS) & \(V_\text{g}\) &  220~V \\
Short circuit ratio & SCR &  4 \\
X/R ratio of the grid & \(n_{XR,\text{g}}\) &  4 \\
DC-link voltage & \(V_\text{dc}\) & 400~V \\
Filter inductance & \(L_\text{f}\) & 3.4~mH \\
Filter capacitance (Y-equivalent) & \(C_\text{f}\) & 30~\(\mu\)F \\
Cutoff frequency for APC and RPC & \(\omega_{p}, \, \omega_{q}\) & 0.1~p.u. \\
Proportional gain for APC and RPC & \(K_{p}, \, K_{q}\) & 0.1~p.u. \\
Control bandwidth of CC & \(\omega_\text{cc}\) & \(2\pi \times\)500~rad/s \\
Virtual inductance & \(L_\text{v}\) & 10~mH \\
Virtual resistance & \(R_\text{v}\) & 0.754~\(\Omega\) \\
Parallel virtual resistance & \(R_{\text{v}\pi}\) & 25.698~\(\Omega\) \\
\hline
\end{tabular}
\end{table}

\begin{figure}[t]
  \centering
  \includegraphics[width=0.8\linewidth,keepaspectratio,trim=0mm 0mm 0mm 0mm,clip]{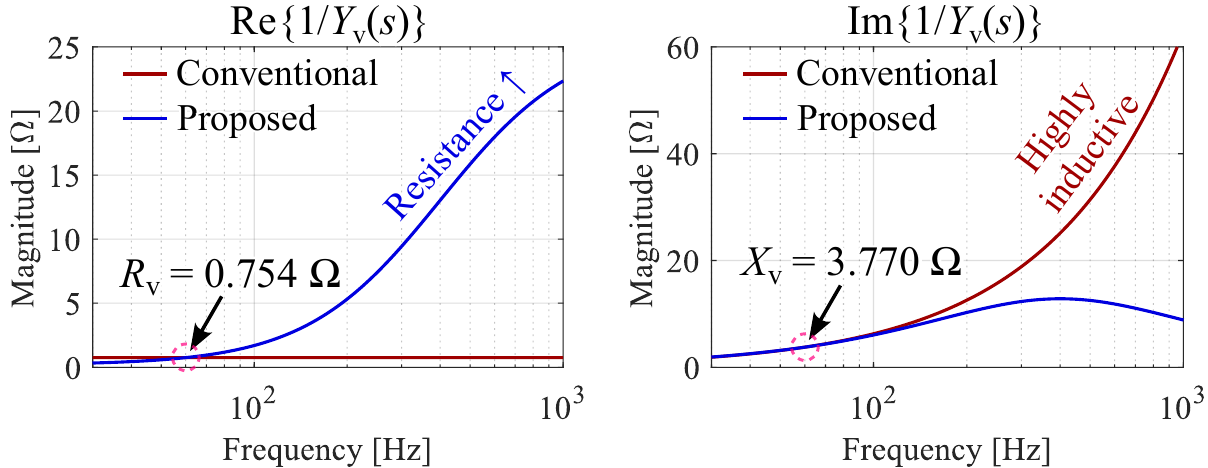}
  \caption{Comparison of \(1/Y_\text{v}(s)\) between the conventional and proposed methods.}
  \label{fig:Zv_compare}
\end{figure}

\begin{figure}[t]
  \centering
  \subfloat[]{
    \includegraphics[width=0.475\linewidth,keepaspectratio,
    trim=0mm 0mm 0mm 1mm,clip]{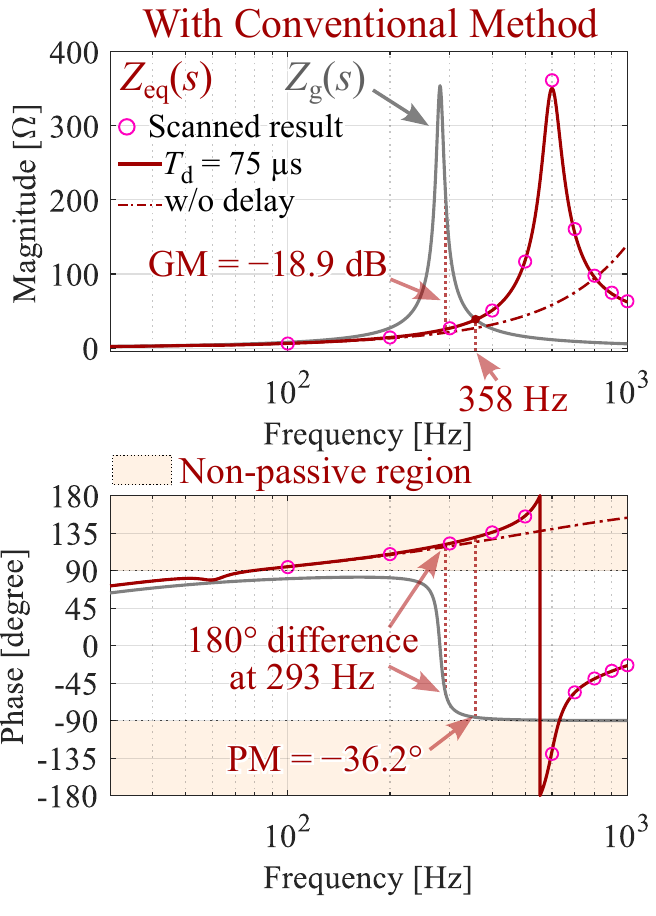}
    \label{fig:ConventionalPlot}
  }
  \subfloat[]{
    \includegraphics[width=0.475\linewidth,keepaspectratio,
    trim=0mm 0mm 0mm 1mm,clip]{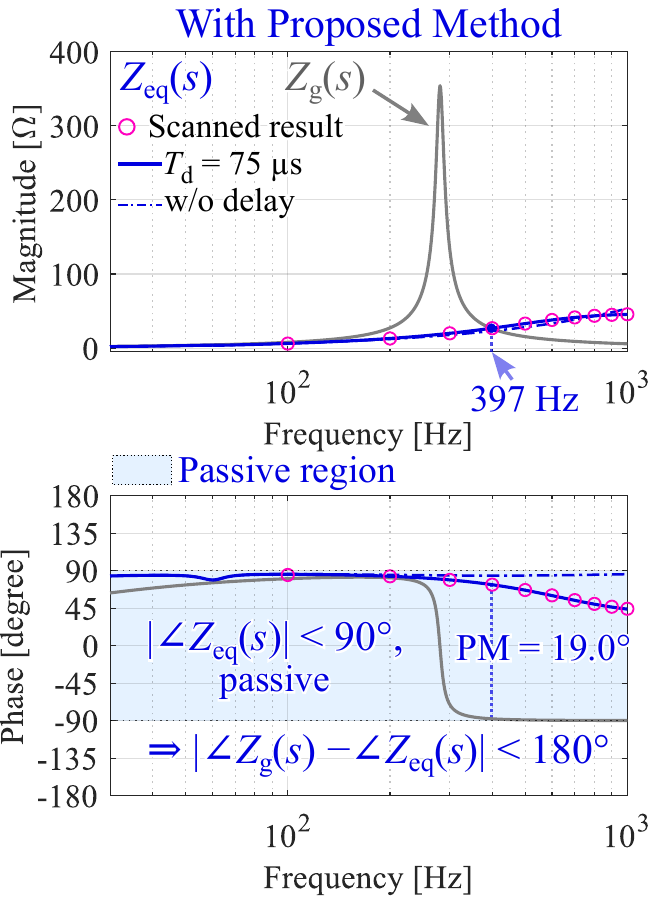}
    \label{fig:ProposedPlot}
  }
  \caption{Impedance plot of \(Z_\text{eq}(s)\): (a) with the conventional VA and (b) with the proposed VA.}
  \label{fig:Conv_vs_Prop}
\end{figure}

A comparative analysis is conducted to evaluate the stability
of the conventional and proposed VA-CC schemes.
The system parameters and configuration used in this study are 
detailed in Table~\ref{tab:parameters} and Fig.~\ref{fig:Setup},
respectively.
The grid-side impedance at the PCC is denoted by \(Z_\text{g}(s)\).
For outer control loop, active power control (APC) and reactive power control (RPC)
are implemented using droop control with low-pass filters \cite{cite8_Rodriguez2013SPC}.
This configuration not only ensures the slow IVS dynamics \cite{cite0_Rosso2021GFM_Review},
but also inherently provides damping against synchronous resonance (SR) \cite{cite9_Zhao2024LowFreqRes}.
Consequently, it obviates the need for a high virtual resistance to prevent SR,
thereby maintaining the high X/R ratio essential for GFM capability \cite{cite5_Wu2024CLC_GFM}.

Fig.~\ref{fig:Zv_compare} compares \(1/Y_\text{v}(s)\) for the
conventional and proposed method.
At the fundamental frequency, two methods are both designed to have \(R_\text{v} = 0.754~\Omega\)
and \(X_\text{v} = 3.770~\Omega\).
However, significant differences appear in the harmonic range,
where the conventional method remains highly inductive,
whereas the proposed method exhibits resistive \(1/Y_\text{v}(s)\).

Using the conventional and proposed \(Y_\text{v}(s)\),
\(Z_\text{eq} (s)\)
is analyzed both with and without control delay.
Considering control delay \(T_\text{d} = 1.5 f_\text{s}^{-1}\) \cite{cite11_Harnefors2016Passivity}, the equivalent impedance is expressed as:
\begin{equation}
  \begin{aligned}
  Z_\text{eq}(s)
  = &
  {e^{-sT_\text{d}} G_\text{cc}(s)}  \left({1 - e^{-sT_\text{d}} + G_\text{cc} (s) Y_\text{v}(s)}\right)^{-1}
  \\
  &+ {sL_\text{f}}  \left( {1 - e^{-sT_\text{d}} + G_\text{cc} (s) Y_\text{v}(s)} \right)^{-1}.
  \end{aligned}
  \label{eq:11}
\end{equation}
Crucially, the structural issue identified in \eqref{eq:3} persists
in the last term of \eqref{eq:11} due to the interaction between
\(sL_\text{f}\) and \(1/Y_\text{v}(s)\).

The frequency response of \(Z_\text{eq}(s)\) with \(Y_\text{v,conv}(s) \) is plotted
in Fig.~\ref{fig:Conv_vs_Prop}(a) for both \(T_\text{d} = 75~\mu\)s and \(0~\mu\)s.
Notably, both cases exhibit non-passive behavior in the harmonic range.
Due to this non-passivity, the stability margin of the return ratio,
\(Z_\text{g}(s)/Z_\text{eq}(s)\), becomes negative.
It is worth mentioning that the both \(T_\text{d}\) cases exhibit similar frequency responses up to the
phase crossover frequency, which occur at 293~Hz
for the case of \(T_\text{d} = 75~\mu\)s.
This indicates that the instability arises from the inherent non-passivity of the conventional VA-CC,
rather than being an artifact of the control delay.

In contrast, with the proposed \(Y_\text{v,prop}(s)\),
the equivalent impedance \(Z_\text{eq}(s)\) exhibits passive behavior
across the entire frequency range, as shown in Fig.~\ref{fig:Conv_vs_Prop}(b).
This ensures that the system remains inherently stable regardless of the grid impedance,
provided that the grid is composed of passive components.
Furthermore, for both the conventional and proposed methods,
the frequency-scanned simulation results closely match the theoretical plots,
thereby validating the proposed analytical framework.

\section{Experimental Verification}

\begin{figure}[t]
  \centering
  \includegraphics[width=0.8\linewidth,keepaspectratio,trim=0mm 0mm 0mm 0mm,clip]{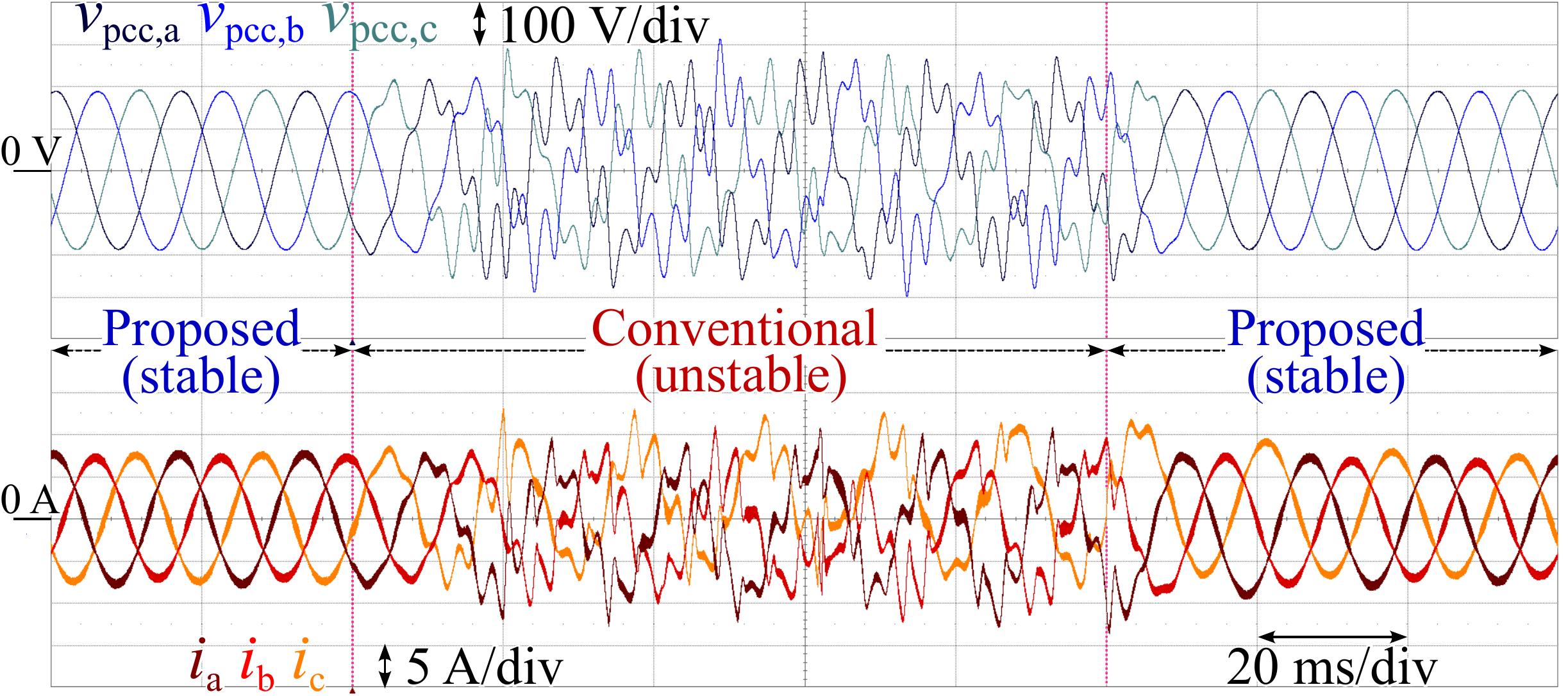}
  \caption{Experimental waveforms under control mode transitions.}
  \label{fig:Experiment_1}
\end{figure}

\begin{figure}[t]
  \centering
  \includegraphics[width=0.8\linewidth,keepaspectratio,trim=0mm 0mm 0mm 5mm,clip]{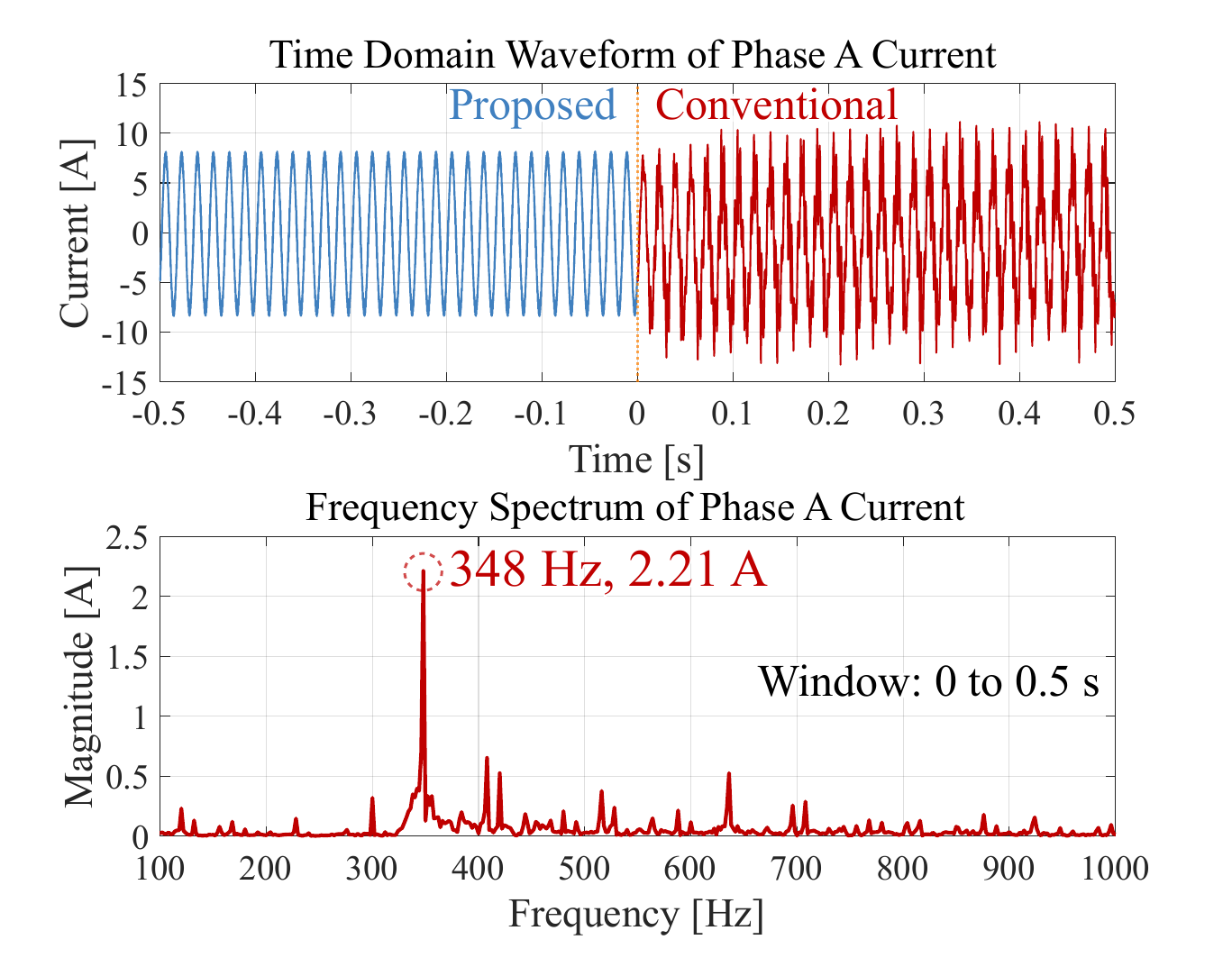}
  \caption{FFT analysis of the phase A current under the conventional method.}
  \label{fig:Experiment_2}
\end{figure}

The experimental results are provided in Fig.~\ref{fig:Experiment_1} and Fig.~\ref{fig:Experiment_2},
with the same conditions summarized in Table~\ref{tab:parameters}.
The reference values are set to
\(P_\text{ref} = 2.0~\text{kW}\) and \(Q_\text{ref} = 0.0~\text{kVAR}\).
This representative operating point is selected to clearly observe the harmonic resonance
of conventional method while avoiding the activation of current-limiting control.
This ensures that the harmonic analysis and FFT result remain unaffected by the
non-linear dynamics of current-limiting functions.
In Fig.~\ref{fig:Experiment_1},
the proposed method with a parallel virtual resistor is initially applied, exhibiting the
stable operation of the GFM inverter.
Subsequently, the conventional VA without the parallel virtual resistor is applied for a duration of 0.1~s.
Under the conventional method, the inverter exhibits harmonic instability as expected.

To further verify the theoretical analysis, FFT results are provided in Fig.~\ref{fig:Experiment_2}.
As shown in the upper plot, the inverter is operated with the proposed VA
as its default configuration, maintaining a sustained steady state.
To observe the harmonic instability, the conventional VA is then activated at 0~s.
The lower plot of Fig.~\ref{fig:Experiment_2} presents the FFT result of the phase~A current,
\(i_\text{a}\), under the conventional method. The analysis reveals a dominant harmonic component at 348~Hz,
which falls within the unstable region identified in Fig.~\ref{fig:Conv_vs_Prop}(a)
and closely aligns with the gain crossover frequency of 358~Hz.

\section{Conclusion}

This letter investigated the inherent instability of VA-CC and provided a
simple yet effective solution by incorporating parallel virtual resistor.
While conventional VA can suffer harmonic instability due to the non-passive
behavior—arising from the interaction between the VA and filter inductance—
the proposed method can ensure a passive output impedance.
This is achieved while preserving the designed X/R ratio and maintaining
the CC and VFF designs.


\begin{thebibliography}{1}

\bibitem{cite0_Rosso2021GFM_Review}
R.~Rosso, X.~Wang, M.~Liserre, X.~Lu, and S.~Engelken,
``Grid-forming converters: Control approaches, grid-synchronization, and future trends—A review,''
\emph{IEEE Open J. Ind. Appl.}, vol.~2, pp.~93--109, 2021,
doi:~10.1109/OJIA.2021.3074028.

\bibitem{cite1_Rodriguez2013VA}
P.~Rodriguez, I.~Candela, C.~Citro, J.~Rocabert, and A.~Luna,
``Control of grid-connected power converters based on a virtual admittance control loop,''
in \emph{Proc. 15th Eur. Conf. Power Electron. Appl. (EPE)}, 2013, pp.~1--10,
doi:~10.1109/EPE.2013.6634621.

\bibitem{cite2_Gao2026VA_MMC}
L.~Gao, J.~Lyu, A.~Shi, X.~Fu, and X.~Cai,
``Small-signal stability analysis and enhancement of virtual admittance control based grid-forming MMC,''
\emph{IEEE Trans. Power Electron.}, vol.~41, no.~4, pp.~5975--5990, 2026,
doi:~10.1109/TPEL.2025.3620354.

\bibitem{cite3_Obi2025RevaluationVA}
S.~A.~Obi and J.-J.~Jung,
``Revaluation of virtual admittance-based dual-loop voltage-controlled grid-forming control considering negative aspects of voltage feedback decoupling on internal stability,''
in \emph{Proc. 10th IEEE Workshop Electron. Grid (eGRID)}, 2025, pp.~1--5,
doi:~10.1109/eGRID63452.2025.11255424.

\bibitem{cite4_Baeckeland2022GFM_FL}
N.~Baeckeland, D.~Venkatramanan, M.~Kleemann, and S.~Dhople,
``Stationary-frame grid-forming inverter control architectures for unbalanced fault-current limiting,''
\emph{IEEE Trans. Energy Convers.}, vol.~37, no.~4, pp.~2813--2825, 2022,
doi:~10.1109/TEC.2022.3203656.

\bibitem{cite5_Wu2024CLC_GFM}
H.~Wu, X.~Wang, and L.~Zhao,
``Design considerations of current-limiting control for grid-forming capability enhancement of VSCs under large grid disturbances,''
\emph{IEEE Trans. Power Electron.}, vol.~39, no.~10, pp.~12081--12085, 2024,
doi:~10.1109/TPEL.2024.3350912.

\bibitem{cite6_Kim1996SkinEffect}
S.~Kim and D.~P.~Neikirk,
``Compact equivalent circuit model for the skin effect,''
in \emph{Proc. IEEE MTT-S Int. Microw. Symp. Dig.}, vol.~3, 1996,
pp.~1815--1818,
doi:~10.1109/MWSYM.1996.512297.

\bibitem{cite7_Matsumori2018LCL_ESR}
H.~Matsumori, T.~Shimizu, F.~Blaabjerg, X.~Wang, and D.~Yang,
``Stability influence of filter components parasitic resistance on LCL-filtered grid converters,''
in \emph{Proc. Int. Power Electron. Conf. (IPEC-Niigata 2018 -- ECCE Asia)}, 2018,
pp.~3357--3362,
doi:~10.23919/IPEC.2018.8507867.


\bibitem{cite8_Rodriguez2013SPC}
P.~Rodriguez, I.~Candela, and A.~Luna,
``Control of PV generation systems using the synchronous power controller,''
in \emph{Proc. IEEE Energy Convers. Congr. Expo. (ECCE)}, 2013,
pp.~993--998,
doi:~10.1109/ECCE.2013.6646811.


\bibitem{cite9_Zhao2024LowFreqRes}
F.~Zhao, T.~Zhu, Z.~Li, and X.~Wang,
``Low-frequency resonances in grid-forming converters: Causes and damping control,''
\emph{IEEE Trans. Power Electron.}, vol.~39, no.~11, pp.~14430--14447, 2024,
doi:~10.1109/TPEL.2024.3424296.

\bibitem{cite11_Harnefors2016Passivity}
L.~Harnefors, X.~Wang, A.~G.~Yepes, and F.~Blaabjerg,
``Passivity-based stability assessment of grid-connected VSCs—An overview,''
\emph{IEEE J. Emerg. Sel. Topics Power Electron.}, vol.~4, no.~1, pp.~116--125, 2016,
doi:~10.1109/JESTPE.2015.2490549.


\end{thebibliography}
\end{document}